\documentstyle[preprint,aps,eqsecnum]{revtex}
\begin{document}

\thispagestyle{empty}
\setcounter{page}{1}

\title{HARD THERMAL LOOPS AND CHIRAL LAGRANGIANS}

\author{Cristina Manuel}

\address{Dpt. Estructura i Constituents de la Mat\`{e}ria\\
Facultat de F\'{\i}sica,
Universitat de Barcelona \\
Diagonal 647,
08028 Barcelona (SPAIN)} 

\maketitle

\thispagestyle{empty}
\setcounter{page}{0}

\begin{abstract}
$\!\!$Chiral symmetry is used as the guiding principle to derive 
hard thermal loop effects in  chiral perturbation theory. 
This is done by using a chiral invariant
background field method for the non-linear sigma model and the
Wess-Zumino-Witten lagrangian, with and without external
vector and axial vector sources. It is then shown that the $n$-point
hard thermal loop is the leading thermal correction for the Green
function of $n$ point vector  soft quark currents.
\end{abstract}

\vfill

\hfill\break
\hbox to \hsize{ECM-UB-PF-97/27} 
\hbox to \hsize{October/1997}
\vskip-12pt
\eject

\baselineskip=15pt
\pagestyle{plain}

\section{INTRODUCTION}
\label{Intro}

At low energies QCD is successfully described in terms of effective 
chiral lagrangians \cite{We,Cole,GassLeut,books}. 
Those only involve the low energy modes or
(pseudo) Goldstone bosons of the spontaneously broken chiral
symmetry.

Chiral symmetry implies that at low energy the functional ${\cal Z}$ 
generating the connected Green's functions of vector and axial vector
quark currents admits a representation of the form
\begin{equation}
\exp{(-i {\cal Z})} = \int [d \Sigma] \exp{( -i \int dx {\cal L}_{eff} (\Sigma, v, a) )} \ ,
\end{equation}
where $\Sigma$ is a $SU(N)$ unitary matrix, $N$ is the number of quark
flavors, and $v_\mu$ and $a_\mu$ are external vector and axial vector sources,
respectively. The effective lagrangian ${\cal L}_{eff}$ is expanded in powers
of external momenta and of quark masses as 
\begin{equation}
{\cal L}_{eff} = {\cal L}_{2} +{\cal L}_{4} +{\cal L}_{6} + \ldots 
\end{equation}
Lorentz invariance, and $P$, $C$ and chiral symmetry restricts the form 
of each term in the series. The matrix $\Sigma$
is parametrized in terms of  pseudoscalar mesons $\pi$ 
as $\Sigma= \exp{(i \pi /f_\pi)}$, where $f_\pi$ can be identified, to
first order, with the pion decay constant, $f_\pi \simeq 93$ MeV.

At low but finite temperature $T$ chiral symmetry is still broken, 
and  the use of effective chiral theory is still valid \cite{GLT}.
This is due to the
fact that in the partition function the contribution of Goldstone states 
dominates over those of heavier particles, which are exponentially
suppressed. The use of chiral effective theory is then justified for
$T \ll f_\pi$, but as  the $T$ is raised,  the contribution of heavier
states becomes important.

Symmetries which are spontaneously broken at zero temperature 
are usually restored at high $T$ \cite{Jackiw}. This is why it is expected that 
chiral symmetry is restored at enough high $T$. 
On the other hand, and due to  asymptotic freedom, at very high $T$
QCD is believed to be
in an unconfined phase \cite{phaseT}, the so called quark-gluon plasma phase. 
Therefore, there are
two different phase transistions in QCD.  It is yet an open question 
whether these occur together or not. Lattice computations seem to indicate that 
the critical temperature of the two phase transitions of QCD are of the same order \cite{lattice}.

At high $T$ and at one-loop a thermal mass for the gauge longitudinal degrees of
freedom  is generated,  the so called Debye mass.  To take into account
properly the thermal effects of Debye screening a resummation of
an infinite number of Feynman diagrams, the hard thermal loops (HTL's),
is required \cite{BP}. Hard thermal loops are thermal amplitudes which arise in
a gauge theory when the external loop
momenta is {\it soft}, while the internal one is {\it hard}.
{\it Soft} denotes a scale $\sim gT$, while {\it hard} refers to one $\sim T$,
where $g \ll 1$ is the gauge coupling constant.

There has been a lot of studies on the HTL's of QCD (see
Ref. \cite{LeBellac} for a 
review).  The effective action generating HTL's  is a mass term for the
chromoelectric fields, and it has been constructed 
just by solving a gauge invariance condition imposed on it \cite{TW}.
This mass term is, however, non-local.

It has been recently claimed that HTL's also arise in the non-linear
sigma model \cite{PT,PT2}. In a recent article Pisarski and Tytgat have shown that
the same thermal amplitudes, the HTL's, for soft modes
 appear in that theory. In this context soft means a energy scale $\ll T$.
In this chiral model HTL's do not describe a mass term, but just thermal corrections
to $\pi - \pi$ scattering.

In this article I present an alternative derivation for the hard thermal effects 
in chiral perturbation theory. The main result of this paper
is showing that chiral symmetry is the guiding principle which allows
to extract the thermal effective action for soft modes in a chiral lagrangian.
This is to be compared with the gauge symmetry principle which was 
imposed in a non-Abelian gauge theory to extract the HTL effective action.

The approach  followed in this paper agrees in spirit with that
of Ref. \cite{PT}, but differs in some technical points. Pisarski and Tytgat
used a background field method to extract the HTL's of a non-linear
sigma model, and the Wess-Zumino-Witten lagrangian. However,
their method was not manifestly chiral invariant.
Here a chiral  invariant background field method will be used instead to find the
symmetry principle which is behind the HTL's of the theory. 
The thermal amplitudes obtained for momenta $P \ll T$ should
agree on-shell for the two different background field methods \cite{PT}.
However, using a manifestly invariant field expansion allows to 
express the one-loop corrections in terms of structures which have
well defined properties under a chiral transformation. This is crucial when
one has to construct higher order lagrangians in chiral perturbation theory.

This paper is structured as follows. In Sec.  \ref{sec2} a derivation of the HTL
effective action in the non-linear sigma model is given. 
In Sec. \ref{sec3} the results are generalized for the non-linear sigma model with
external vector and axial vector sources. The method of external sources
allows to recognize that the $n$-point HTL, with $n = 2, \ldots, \infty$,
is the leading thermal correction for the Green function of $n$ soft vector
quark currents. In Sec. \ref{sec4}  the Wess-Zumino-Witten terms, with and
without external
sources, are taken into consideration, and the hard thermal effects are
studied. Once again, a symmetry 
principle allows to get the corresponding 
thermal effective action.  Conclusions
are presented in the last section. Throughout this article
computations are done using the 
imaginary time formalism, and it is assumed that $T \ll f_\pi$.

\section{THE NON-LINEAR SIGMA MODEL}
\label{sec2}

Let us  consider the lagrangian of the non-linear sigma model

\begin{equation}
{\cal L}  = \frac{f_\pi^2}{4} Tr \left (\partial_\mu \Sigma^{\dagger} \partial_\mu 
\Sigma
\right) \ , \qquad \Sigma^{\dagger} \Sigma = 1 
\label{lagnonsi}
\end{equation}
where $\Sigma$ is a $SU(N)$ unitary  matrix. The above lagrangian is 
invariant under a global  $SU_R (N) \times SU_L (N)$ symmetry, where
$\Sigma$  transforms as  
$\Sigma' (x) = U_R \Sigma (x) U_L  ^{\dagger}$,  
and $U_{R,L} \in SU_{R,L} (N)$.

A standard technique to evaluate the loop effects generated by the
lagrangian ${\cal L}$ consists in expanding it around the 
solution of the classical equations of motion. 
The one-loop effective action is then obtained after integrating out the
quantum fluctuations. There are several different ways to perform the
expansion around the classical solution. In this section a background field 
expansion respectful with the symmetries of ${\cal L}$ will be performed
to compute one-loop thermal effects.

Let $\bar{\Sigma}$ be the solution of the classical  equations of motion
derived from  (\ref{lagnonsi}). One now defines the field $\xi(x)$ as
the square root of $\bar{\Sigma}$ \cite{GassLeut,books}
\begin{equation}
\bar{\Sigma} (x) = \xi(x) \xi(x) \ .
\end{equation}

The field $\Sigma$ is then written  in the form
\begin{equation}
\Sigma (x) = \xi(x) h(x) \xi(x)  \ .
\label{split}
\end{equation}

It is interesting to study  the transformation properties of
these new variables. Under a global $SU_R (N) \times SU_L (N)$
transformation the field $\xi(x)$ behaves as
\begin{equation}
\xi'(x) = U_R\, \xi (x) U^{\dagger} (x) = U(x) \xi(x)   U_L  ^{\dagger} \ ,
\label{compen1}
\end{equation} 
while 
\begin{equation}
h'(x) = U(x) h(x) U^\dagger (x) \ ,
\label{compen2}
\end{equation}
where $U$ is a unitary matrix which depends on $\xi(x)$ (and thus it
depends on $x$!), $U_R$ and 
$ U_L  ^{\dagger}$. If $ U_R = U_L$,  then $U= U_R = U_L$ and then
$U$ is associated to an ordinary  $SU(N)$ transformation.

Let us write down the lagrangian (\ref{lagnonsi}) in terms of the 
new variables $\xi$ and $h$. It is not difficult to check 
\begin{eqnarray}
{\cal L}  & = & \frac{f_\pi^2}{4} Tr \left (\partial_\mu \left(\xi h
 \xi \right)^\dagger \partial_\mu \left(\xi h \xi \right) \right) \nonumber \\ 
& = & \frac{f_\pi^2}{4}  Tr \left( D_\mu h^\dagger D_\mu h \right) \ ,
\label{lagqusi}
\end{eqnarray}
where
\begin{eqnarray}
D_\mu h & = & \partial_\mu h + L_\mu h -  h R_\mu \ ,   \\
D_\mu h^\dagger  & = & \partial_\mu h^\dagger - h^\dagger L_\mu + R_\mu
h^\dagger \ ,
\end{eqnarray}
with $L_\mu$ and $R_\mu$ being defined as
\begin{equation}
L_\mu = \xi^\dagger  \partial_\mu \xi  \ , \qquad  
R_\mu =  \xi \partial_\mu \xi^\dagger \ .
\end{equation}

One can deduce from (\ref{compen1}) the transformation properties of the above fields
under the global  $SU_R (N) \times SU_L (N)$ symmetry.
These read
\begin{mathletters}
\begin{eqnarray}
R' _\mu (x) & =& U (x) R _\mu (x) U^\dagger (x) + U (x) \partial_\mu 
 U^\dagger (x)  \ ,
\label{setru1}
 \\
L' _\mu (x) & =& U (x) L _\mu (x) U^\dagger (x) + U (x) \partial_\mu
 U^\dagger (x) \ .
\label{setru2}
\end{eqnarray}
\end{mathletters}

At this point, one parametrizes  $h = \exp(i \pi/f_\pi)$, where 
$\pi$ is an hermitian and traceless matrix. Then  ${\cal L}$ is expanded
keeping only terms up and including quadratic in $\pi$. The linear term in $\pi$
vanishes after imposing the classical equations of motion.
One then gets
\begin{equation}
{\cal L}^{(0)} + {\cal L}^{(2)} =  - f_\pi^2 \, Tr  (A_\mu ^2)
+ \frac 14 \,  Tr \left( D_\mu \pi \right)^2 - \frac14 \, Tr ([A_\mu ,\pi])^2 \ ,
\label{lagpi}
\end{equation}
where
\begin{equation}
D_\mu \pi  =  \partial_\mu \pi +  [V_\mu, \pi] \ ,
\end{equation}
\begin{equation}
V_\mu (x)  =   \frac{L_\mu (x) + R_\mu (x)}{2}  \ ,  \qquad
A_\mu (x)  =  \frac{ L_\mu (x) - R_\mu (x)}{2}   \ .
\end{equation}

It is possible to deduce from (\ref{compen2}) and (\ref{setru1})-(\ref{setru2})
how the fields $\pi$, $V_\mu$ and $A_\mu$ behave under a chiral transformation.
Thus
\begin{mathletters}
\begin{eqnarray}
\label{compset}
\pi' (x) & = & U (x) \pi(x) U^\dagger (x)  \ , 
\label{chang1}\\
V' _\mu (x) & =& U (x) V _\mu (x) U^\dagger (x) + U (x) \partial_\mu
 U^\dagger (x) \ , \label{chang2}\\
A' _\mu (x) & =& U (x) A_\mu (x) U^\dagger (x) \label{chang3} \ .
\end{eqnarray}
\end{mathletters}

The field $V_\mu$ transforms like a connection, while $ A_\mu$ and $\pi$
transforms
covariantly. Thus, it can be immediately checked that  under the above symmetry
each term in (\ref{lagpi}) remains invariant.
The lagrangian (\ref{lagpi}) looks formally as the one of a non-Abelian gauge
theory, $V _\mu$ being the  corresponding vector gauge field. This
similarity is just formal, since there is not a kinetic term for 
$V _\mu$, neither for  $A_\mu$, and thus those fields do not propagate.

The one-loop effective action of the non-linear sigma model is
obtained by integrating out the $\pi$ fields. At zero temperature
it can be done by evaluating the determinant of a differential operator,
since the action is quadratic in the $\pi$ fields\footnote{Note also that the jacobian
of the change of variables is unit at one-loop order.}.

It is our aim here to compute the leading thermal corrections
to the effective action. The UV divergences which appear at finite $T$
are actually the same as the ones at $T=0$. 
The $T=0$ contributions to the effective action and the 
renormalization of the model will not be explicitly taken into account here.
For  the $T=0$ results see Ref. \cite{GassLeut}.

At this point it is interesting to make contact with the analysis
of non-Abelian gauge theories at finite temperature. As pointed out
by Pisarski and Tytgat, HTL's arise also in this model 
when the momenta of the background fields are soft, {\rm i.e.} 
$\ll T$. 

Let us first compute the thermal two-point functions
 $\langle V_\mu ^a V_\nu ^b \rangle_T$
and $\langle A_\mu ^a A_\nu ^b\rangle_T$. 
The computation of these functions is done here 
using the imaginary time formalism (ITF), following the techniques
developed in Ref. \cite{BP}.

In terms of the components $\pi^a$, we write $\pi = \pi^a \lambda^a$, where
$\lambda^a$ is a generator of $SU(N)$ normalized as 
$Tr (\lambda^a \lambda^b) = 2 \delta^{ab}$, and $[\lambda^a, \lambda^b] =
2 i f^{abc}\lambda^c$, where $f^{abc}$ are the structure constants of
$SU(N)$.
The $V_\mu$ two-point function
then reads
\begin{equation}
C_{V, \mu \nu}^{ab} (P) =  f^{adc} f^{bdc} \left(- 2 \, \delta_{\mu \nu}
 \int  \frac{d^4 K}{(2 \pi)^4}
\frac{1}{(K+P)^2} 
 +  \int \frac{d^4 K}{(2 \pi)^4} \frac{(2 K + P)_\mu (2 K + P)_\nu}
{K^2 (K+P)^2} \right) \ .
\label{twoVp}
\end{equation}

In ITF $K^2 = k_0 ^2 + {\bf k}^2$, and for bosonic fields $k_0 = 2\pi nT$
for integral $n$.  The standard notation
\begin{equation}
\int  \frac{d^4 K}{(2 \pi)^4} \equiv T \sum_{n=-\infty}^{n=\infty}
\int  \frac{d^3 k}{(2 \pi)^3} \ , 
\end{equation}
is used throughout this article.

The leading correction in  $T$ when the external momentum 
$P$ is soft, i.e. $\ll T$, arises when the internal one $K$  is hard, i.e. $\sim
T$. In this case, one can neglect  $P$ in front
of $K$ in the numerator of (\ref{twoVp}), and then one gets
\begin{equation}
C_{V, \mu \nu}^{ab} (P)  \approx  -2 N \delta^{ab}  
  \int \frac{d^4 K}{(2 \pi)^4} \frac{(K^2
 \delta_{\mu \nu} -2 K _\mu  K _\nu)} {K^2 (K+P)^2} \ .
\label{add2}
\end{equation}
The $SU(N)$ relation $f^{adc} f^{bdc} = N \delta^{ab}$ has been used
in Eq. (\ref{add2}).

It is possible to split the pure $T=0$ part from the finite
$T$ contribution of the two-point function.
 We will only concentrate in the last one. Its dominant term can be extracted
in the limit where the external momentum is soft,
and it reads
\begin{equation}
C_{V, \mu \nu}^{ab} (P, T) \approx - \delta^{ab} \frac{N T^2}{6}
\delta \Pi_{\mu \nu} (P) \ ,
\label{htlpol}
\end{equation}
where $\delta \Pi_{\mu \nu} (P)$ is the HTL arising in the
polarization tensor of a gauge field theory \cite{BP}. This can be expressed as

\begin{equation}
\delta \Pi_{\mu \nu} (P) = 2 \delta_{\mu 0}\delta_{\nu 0} + 2 i p_0 
\, \int \frac{d \Omega_{\bf q}}{4 \pi} \, \frac{Q_\mu Q_\nu}{Q \cdot P} \ ,
\end{equation}
where $Q = (i, {\bf q})$ is a null vector, $Q^2 =0$, and the angular integral
is over all directions of the three dimensional unit vector ${\bf q}$.

The thermal $A_\mu$ two-point function can be evaluated following a 
similar analysis. One then gets
\begin{equation}
\label{AtwoT}
C_{A, \mu \nu}^{ab} (P, T) \approx  \delta^{ab}
 \delta_{\mu \nu} \, \frac{N T^2}{6} \ ,
\end{equation}
and the $\approx$ sign means that only the leading thermal correction is
retained.

Although the thermal correction  in (\ref{AtwoT})
gives a contribution
to the effective  action which is invariant under the symmetry
(\ref{chang1})-(\ref{chang3}), the thermal amplitude
$Tr(V_\mu \delta \Pi^{\mu \nu} V_\nu)$ does not. 
Since  $V_\mu$ transforms as a connection under a chiral transformation
it cannot appear in a chiral invariant  effective action. However, 
as  known from the studies of non-Abelian gauge theories, 
this is not the only thermal amplitude contributing to the effective
action. There are also  higher $n$-point functions, with $n$ arbitrary, which 
are relevant at the same order in the computation.
Those can be determined by an explicit evaluation of each Feynman diagram
\cite{BP}.

In a gauge theory it has been proved that in the HTL approximation 
the quartic couplings of (\ref{lagpi}) only contribute
in tadpole-like diagrams, but not in higher $n$-point functions.
Therefore in the non-linear sigma model the only HTL involving the $A_{\mu}$ 
fields are the two-point functions. This is also the reason why there are 
no mixed $A_{\mu}-V_{\mu}$ thermal amplitudes in this approximation.
There is, however, an infinite set of 
HTL's involving the $V_{\mu}$ fields.

A symmetry argument allows to extract the complete thermal
effective action for soft background fields, which is then expressed in terms 
of covariant objects. Following closely the work done in a gauge theory
\cite{BP2}, one can write the effective action as

\begin{eqnarray}
S + \delta S_{T} & = & -f_{\pi}^2 (T) \int d^4 x \, Tr (A^2_\mu (x))
\nonumber \\
&- &\frac{N T^2}{12}\int \frac{d \Omega_{\bf q}}{4 \pi} \int d^4 x \,d^4 y\,
Tr \left(V_{\mu \lambda} (x) <x | \frac{Q^\mu Q_{\nu}}{-(Q \cdot
D)^2} |y> V^{\nu \lambda} (y) \right) \ ,
\label{HTLeffac}
\end{eqnarray}
where 
\begin{equation}
V_{\mu \nu} = \partial_\mu V_\nu -  \partial_\nu V_\mu
+  [V_\mu, V_\nu]  \ .
\end{equation}

Hard thermal effects change $f_\pi$ into $f_\pi(T)$ as
\begin{equation}
f_{\pi} (T) = f_\pi \left( 1 - \frac{N}{24} \frac{T^2} {f_\pi ^2} \right)
\end{equation}
which agrees with the result computed in the literature (see Ref. 
\cite{GLT,fT}).

Notice that although the field strentgh $V_{\mu \nu}$ did not
show up in ${\cal L}$, it is generated by the thermal
corrections. In a real gauge field theory the second piece
in (\ref{HTLeffac}), which is non-local,
 represents a gauge invariant mass term for the 
longitudinal degrees of freedom.
 In the non-linear sigma model,
and after writing down  $V_\mu$  in terms of the $\pi$ fields, 
the same HTL  action  represents thermal $\pi -\pi$
scattering \cite{PT} respectful with the symmetries of the
theory.

In the static limit the non-local term in (\ref{HTLeffac}) becomes local and
proportional to $Tr(V_0^2)$. However, since in the static situation $V_0 =A_0 =0$,
then the only hard thermal correction in (\ref{HTLeffac})
 which survives in that case is
$-f_\pi ^2 (T) \, \int dx \, Tr(A_i^2)$. 

At this point it is interesting to compare our results with the ones obtained by
Pisarski and Tytgat. Those authors performed a different background field
expansion to get the  hard thermal effects.  With some minor modifications
in the notations, instead of the splitting (\ref{split}),
Pisarski and Tytgat defined
$\Sigma (x) = {\bar \Sigma}(x) h (x)$, ${\bar \Sigma}$ being the
background field, and $h= \exp{(i \pi /f_\pi)}$,
$\pi$ being the quantum fluctuation.  Written in
terms of the $h$ fields, one then gets a lagrangian which looks formally
the same as  (\ref{lagqusi}), but with $R_\mu = 0$ and ${\tilde L}_\mu = 
{\bar \Sigma}^{\dagger} \partial_\mu {\bar \Sigma}$.
After expanding $h = \exp{(i \pi/f_\pi)}$, and keeping only the terms
up to quadratic in the $\pi$ fields, one then gets
\begin{equation}
{\cal L}^{(0)} +{\cal L}^{(2)}   = - \frac{f^2_\pi}{4} Tr  ({\tilde L}_\mu ^2)
+ \frac 14 \,  Tr \left( {\tilde D}_\mu \pi \right)^2 -\frac{1}{16} \, Tr
 ([{\tilde L}_\mu ,\pi])^2 \ ,
\label{lagPT}
\end{equation}
with ${\tilde D}_\mu \pi = \partial_\mu \pi + \frac 12 [{\tilde L}_\mu, \pi]$.
This lagrangian is not  invariant under a standard
$SU(N)$ transformation, since both the first and third terms  
break the symmetry. 
Nevertheless, Pisarski and Tytgat invoked a $SU(N)$ gauge symmetry principle to
write the complete one-loop thermal effective action. According to
Ref. \cite{PT}, this 
was found to be 
\begin{eqnarray}
{\tilde S} + \delta {\tilde S}_{T} & = & -\frac{f_{\pi}^2 (T)}{4} \int d^4 x \, Tr (
{\tilde L}^2_\mu (x)) \nonumber
 \\
&- &\frac{N T^2}{12}\int \frac{d \Omega_q}{4 \pi} \int d^4 x \,d^4 y\,
Tr \left({\tilde L}_{\mu \lambda} (x) <x | \frac{Q^\mu
Q_{\nu}}{-(Q \cdot {\tilde D})^2} |y> {\tilde L}^{\nu \lambda} (y) \right) \ ,
\label{PTlag}
\end{eqnarray}
where 
\begin{equation}
{\tilde L}_{\mu \nu} = \partial_\mu {\tilde L}_\nu -  \partial_\nu {\tilde L}_\mu
+ \frac 12 [{\tilde L}_\mu, {\tilde L}_\nu]  \ .
\end{equation}
Notice that the first term is not invariant under a gauge $SU(N)$ symmetry,
while the second is by construction.

At first sight, Eqs. (\ref{HTLeffac}) and (\ref{PTlag}) seem to be different.
However, they should give the same thermal amplitudes on-shell \cite{PT}.
To check agreement between the two formalisms, use of the Ward
identities obeyed by the HTL's is required \cite{WId}.
Here only agreement between the two-point functions will be
checked in the lowest order in $1/f_\pi$. 

After writing $\xi = \exp{ (i \pi/2 f_\pi)}$, and expanding 
the exponentials, the $A_\mu$ and $V_\mu$ fields are 
expressed as
\begin{equation}
A_\mu = \frac{i}{2 f_\pi} \partial_\mu \pi + O (\frac{1}{f_\pi ^3})
 \ , \qquad V_\mu = \frac{1}{8 f_\pi ^2} [\pi, \partial_\mu \pi]
+ O (\frac{1}{f_\pi ^4}) \ .
\end{equation}

Following Ref. \cite{PT} ${\tilde L}_{\mu}$ is given by
\begin{equation}
{\tilde L}_\mu = \frac{i}{f_\pi} \partial_\mu \pi + 
\frac{1}{2 f_\pi ^2} [\pi, \partial_\mu \pi] +
 O (\frac{1}{f_\pi ^3}) \ .
\label{acuer}
\end{equation}

Since the  HTL obeys the 
Ward identity $P_\mu \delta \Pi^{\mu \nu} (P) = 0$, 
the contribution of the first term of (\ref{acuer})
vanishes in the two-point function 
$Tr ({\tilde L}_\mu \delta \Pi^{\mu \nu}{\tilde L}_\nu)$. I then found agreement
between the two formalisms on-shell, except for a factor of 4, which I think it was
introduced incorrectly in (\ref{PTlag}) in Ref. \cite{PT}.

\section{ THE NON-LINEAR SIGMA MODEL WITH EXTERNAL SOURCES}
\label{sec3}

In this section we consider the non-linear sigma model
\begin{equation}
{\cal L}_2  = \frac{f_\pi^2}{4}\left( Tr \left(\nabla_\mu \Sigma^{\dagger} \nabla_\mu 
\Sigma\right) + Tr \left( \chi^{\dagger} \Sigma + \chi  \Sigma^{\dagger} \right)\right)
 \ , \qquad \Sigma^{\dagger} \Sigma = 1 \ ,
\label{lagnonsiga}
\end{equation}
where the covariant derivatives are defined as
\begin{equation}
\nabla_\mu \Sigma = \partial_\mu \Sigma - i (v_\mu + a_\mu) \Sigma + 
i  \Sigma (v_\mu - a_\mu) \ ,
\end{equation}
$v_\mu$ and $a_\mu$ being external vector and axial vector 
sources, respectively, and $\chi = B (s+i p)$, where $B$ is a
constant, and $s$ and $p$ are scalar and pseudoscalar external sources,
respectively.

It is convenient to introduce  the combinations
\begin{equation}
F_\mu ^R = v_\mu + a_\mu \ , \qquad F_\mu ^L = v_\mu - a_\mu \ .
\end{equation}

Electromagnetic interactions in the non-linear sigma model can be considered
as a special case of (\ref{lagnonsiga}), just by setting $a_\mu =0$, and
$v_\mu = F_\mu^R =F_\mu^L = Q A_\mu ^{e.m.}$, where $A_\mu ^{e.m.}$ is the
electromagnetic gauge field, and $Q$ is the quark charge matrix, which
for the $N=3$ case reads
\begin{equation}
Q = \frac e3 \pmatrix{
2 &  &  \cr
 & -1 &   \cr
 &    & -1 \cr} \ .
\end{equation}

Under the local $SU_R (N) \times SU_L (N)$ symmetry the fields 
transform as 
\begin{mathletters}
\begin{eqnarray}
\Sigma' (x) & = & U_R (x) \Sigma (x) U_L  ^{\dagger} (x) \ , \\ 
F'^R _\mu (x) & =& U_R (x)  F^R _\mu (x) U_R^\dagger (x) +
 i U_R (x) \partial_\mu
 U_R^\dagger (x) \\
F'^L _\mu (x) & =& U_L (x) F^L _\mu (x) U_L^\dagger (x) + i U_L (x) \partial_\mu
 U_L^\dagger (x) \ , \\
(s'(x) + i p'(x)) & = & U_R (x) (s(x) + i p(x)) U_L  ^{\dagger} (x) \ .
\end{eqnarray}
\end{mathletters}

The same analysis as in the previous section will be followed to
compute the leading thermal corrections of the one-loop effective action,
working in the exact chiral limit, so that $s=p=0$. In this limit
the classical equations of motion derived from (\ref{lagnonsiga})
are
\begin{equation}
\bar{\Sigma} \nabla_\mu \nabla_\mu \bar{\Sigma}^\dagger -  \nabla_\mu \nabla_\mu
\bar{\Sigma} \bar{\Sigma}^\dagger = 0 \ .
\end{equation}
As in Sec. \ref{sec2} , one then defines
\begin{equation}
\Sigma (x) = \xi(x) h(x) \xi(x)  \ ,
\end{equation}
where  $\bar{\Sigma} = \xi^2$ is the classical solution to the equations of motion.
The transformation properties of $\xi$ and $h$ under the  local 
$SU_R (N) \times SU_L (N)$ symmetry are
\begin{mathletters}
\begin{eqnarray}
\xi'(x) & = & U_R(x) \, \xi (x) U^{\dagger} (x) = U(x) \xi(x)   U_L  ^{\dagger} (x) \ ,
\label{compenga1} \\
h'(x) & = & U(x) h(x) U^\dagger (x) \ ,
\label{compenga2}
\end{eqnarray}
\end{mathletters}
where $U$ is a unitary matrix which depends on $\xi(x)$, $U_R(x)$ and
 $U_L  ^{\dagger} (x)$.

Then one  writes $h=\exp{(i \pi/f_\pi)}$ and expands the exponentials, keeping only terms up and including 
the quadratic in $\pi$ in the lagrangian. Thus 
\begin{equation}
{\cal L}_2 ^{(0)} + {\cal L}_2 ^{(2)} = 
{\cal L}_2 ({\bar \Sigma}) + \frac{1}{4}  
Tr \left(\nabla_\mu (\xi^\dagger \pi \xi^\dagger)
\nabla_\mu (\xi \pi \xi) - \frac 12 \nabla_\mu {\bar \Sigma}^\dagger \nabla_\mu
(\xi \pi^2  \xi) -\frac 12 \nabla_\mu {\bar \Sigma} 
\nabla_\mu (\xi^\dagger \pi^2 \xi^\dagger) \right) \ .
\end{equation}

It is possible to express the above lagrangian as \cite{GassLeut}
\begin{equation}
\label{anlag}
{\cal L}_2 ^{(0)} + {\cal L}_2 ^{(2)}   = - f^2_\pi \, Tr (\Delta_\mu)^2 +
\frac{1}{4} Tr (d_\mu \pi)^2 - \frac{1}{4} Tr ([\Delta_\mu, \pi])^2 \ ,
\end{equation}
where 
\begin{mathletters}
\begin{eqnarray}
d_\mu \pi& = &  \partial_\mu \pi +  [ \Gamma_\mu, \pi] \ , \\
\Gamma_\mu & = & \frac 12 \left( \xi^\dagger \nabla_\mu ^R \xi + \xi \nabla_\mu ^L 
\xi^\dagger \right) \ , \\
\Delta_\mu & = &  \frac 12 \left( \xi^\dagger \nabla_\mu ^R \xi - \xi \nabla_\mu ^L 
\xi^\dagger \right) \ , \\
\nabla_\mu ^l & = & \partial_\mu - i F_\mu ^l \ , \qquad l = R, L  \ .
\end{eqnarray}
\end{mathletters}

The transformation rules obeyed by the new  fields are then deduced from
(\ref{compenga1}) - (\ref{compenga2})
\begin{mathletters}
\begin{eqnarray}
\pi' (x) & = & U (x) \pi(x) U^\dagger (x) \ ,  \\
\Gamma' _\mu (x) & =& U (x) \Gamma _\mu (x) U^\dagger (x) + U (x) \partial_\mu
 U^\dagger (x) \ , \\
\Delta' _\mu (x) & =& U (x) \Delta_\mu (x) U^\dagger (x) \ .
\end{eqnarray}
\end{mathletters}

The lagrangian (\ref{anlag})  looks like the same as the 
one in (\ref{lagpi}), 
with $\Gamma_\mu$ and $\Delta_\mu$ playing the same role as 
the $V_\mu$ and $A_\mu$ fields, respectively.

The same analysis as the one performed in the previous section leads to the
thermal one-loop effective action for soft background fields
$\Gamma_\mu$ and $\Delta_\mu$. It reads
\begin{eqnarray}
{\cal Z}_2 = S_2 + \delta S_{2,T} & = & - f_{\pi}^2 (T) \int d^4 x \, Tr (\Delta^2_\mu (x))
\nonumber \\
&- &\frac{N T^2}{12}\int \frac{d \Omega_{\bf q}}{4 \pi} \int d^4 x \,d^4 y\,
Tr \left(\Gamma_{\mu \lambda} (x) <x | \frac{Q^\mu Q_{\nu}}{- (Q \cdot
d)^2} |y> \Gamma^{\nu \lambda} (y) \right) \ ,
\end{eqnarray}
where 
\begin{equation}
\Gamma_{\mu \nu} = \partial_\mu \Gamma_\nu -  \partial_\nu \Gamma_\mu
+  [\Gamma_\mu, \Gamma_\nu]  \ .
\end{equation}

For later convenience,  an identity obeyed by $\Gamma_{\mu \nu} $ is given
\cite{GassLeut}
\begin{equation}
\Gamma_{\mu \nu}  = - [\Delta_\mu, \Delta_\nu] - \frac 12 F_{\mu \nu} ^{+} \ ,
\label{useid}
\end{equation}
where 
\begin{equation}
F_{\mu \nu} ^{\pm} = i \left(\xi F_{\mu \nu} ^L \xi^{\dagger} \pm \xi^{\dagger} 
F_{\mu \nu} ^R \xi \right) \ .
\end{equation}

Chiral symmetry implies that the Green's functions associated
with the vector and axial vector quark currents
\begin{equation}
J_\mu ^a (x) = \bar{q} (x) \frac{\lambda^a}{2} \gamma_\mu q(x) \
, \qquad J_{\mu 5} ^a (x) = \bar{q} (x) \frac{\lambda^a}{2}
 \gamma_\mu \gamma_5 q(x) \ ,
\end{equation}
can be obtained in the lowest order in the external momenta
from the generating functional ${\cal Z}_2$. This can be done
by performing functional derivatives
of  ${\cal Z}_2$ with respect to the external sources.
Thus
\begin{eqnarray}
\Big. \frac{\delta^2 {\cal Z}_2}{\delta a_\mu ^a (x) \delta a_\nu ^b
(y)} \Big |_{a=v=0}&  = &\langle  J_{\mu 5} ^a (x)  J_{\nu 5} ^b (y) \rangle \ ,
\\
\Big. \frac{\delta^2 {\cal Z}_2}{\delta v_\mu ^a (x) \delta
v_\nu ^b
(y)} \Big |_{a=v=0}&  = & \langle J_{\mu } ^a (x)  J_{\nu } ^b (y) \rangle \ .
\end{eqnarray}

Thermal effects modify the above two Green´s functions with
respect to their $T=0$ values \cite{GLT,fT}.
Furthermore, the HTL effective 
action gives account of the leading thermal $n$-point vector quark current amplitudes.
In momentum space, and for $P_i \ll T$, one has

\begin{equation}
\langle J_{\mu_1 } ^{a_1} (P_1) \ldots J_{\mu_n } ^{a_n} (P_n)\rangle_T \approx
(-i)^n \frac{N T^2}{12} \delta \Gamma ^{a_1 \ldots a_n} _{\mu_1 \ldots \mu_n}
(P_1,\ldots , P_n) + O (\frac{1}{f_\pi ^2}) \ ,
\end{equation}
where $\delta \Gamma ^{a_1 \ldots  a_n} _{\mu_1 \ldots  \mu_n}
(P_1, \ldots, P_n) $ is the $n$-point HTL.

\section{THE WESS-ZUMINO-WITTEN LAGRANGIAN}
\label{sec4}

To incorporate correctly the anomalies of QCD in presence of external
vector and axial vector fields one has also to include in the chiral lagrangian
the Wess-Zumino terms \cite{WZW}.
In chiral perturbation theory those terms are of order $P^4$.
 In this section  the leading thermal corrections
generated by those new terms will be computed.
Computations at finite $T$  with the Wess-Zumino-Witten (WZW) lagrangian
have already been considered in Ref. \cite{AE,PT,PT2}.

The Wess-Zumino-Witten action in the absence of external sources can be expressed
as an integral in five dimensions as \cite{WZW}
\begin{equation}
\label{WZaction}
S_{WZ} (\Sigma) = - \frac{i N_c}{240 \pi^2} \int_{\partial D= M^4} 
\epsilon^{\alpha \beta \gamma \delta \rho}\,
 Tr \left(\Sigma^{-1} \partial_{\alpha} \, \Sigma
\Sigma^{-1} \partial_{\beta} \Sigma \, \Sigma^{-1} \partial_{\gamma} \Sigma
\,\Sigma^{-1} \partial_{\delta} \Sigma \, \Sigma^{-1} \partial_{\rho} \Sigma \right) d^5 x
 \ .
\end{equation}
In Eq. (\ref{WZaction}) $N_c$ is the number of colors, and the integral is over a
five dimensional surface $D$ whose boundary is the four dimensional space.

In the presence of external vector and axial vector gauge fields the 
Wess-Zumino-Witten action is given by \cite{AGG}
\begin{eqnarray}
\label{WZWaction}
S_{WZ} (F^L, F^R, \Sigma) & = & - \frac {i N_c} {240 \pi^2} \int_{\partial D= M^4}
Tr \left (\Sigma^{-1} d \Sigma \right)^5 \\
& - &  \frac {N_c} {48 \pi^2} \int Tr \left( -i F^R (d \Sigma \Sigma^{-1})^3 - P.C.
\right.
\nonumber \\
& + &  \left( -F^R dF^R - dF^R F^R + i (F^R)^3  \right)
 \left(-i \Sigma F^L \Sigma^{-1} + \Sigma^{-1} d \Sigma \right) - P.C. 
\nonumber \\
& -&  \left( d \Sigma \Sigma^{-1} dF^R \Sigma F^L \Sigma^{-1} - P.C. \right)
\nonumber \\
&-& \frac 12  \left(F^R d \Sigma  \Sigma^{-1} F^R d \Sigma \Sigma^{-1} -P.C. 
 \right)
\nonumber \\
&-&  \left(\Sigma F^L \Sigma^{-1} F^R d \Sigma \Sigma^{-1}
 d \Sigma \Sigma^{-1} - P.C.  \right)
\nonumber \\
&-& i \left(F^L \Sigma^{-1} d \Sigma F^L \Sigma^{-1} F^R \Sigma - P.C. \right)
\nonumber \\
&-&  \left. \frac 12  \left(F^L \Sigma^{-1} F^R  \Sigma F^L \Sigma^{-1} F^R \Sigma
\right) \right) \ , 
\nonumber
\end{eqnarray}
where $P.C.$ stands for parity conjugate:
$F^L \leftrightarrow F^R$ , $  \Sigma  \leftrightarrow \Sigma^{-1}$.
The language of differential forms has been used in (\ref{WZWaction}), so
$d \Sigma  = \partial_\mu \Sigma\, d x^\mu$, $F^{L,R} =F^{L,R}_\mu d x^\mu$, etc.

A background field expansion of the total lagrangian
${\cal L}_2 + {\cal L}_{WZ}$ will be performed   
keeping only the quadratic terms  in the fluctuations.
In order to carry out this expansion it is very convenient
to use the techniques advocated by Akhoury and Alfakih \cite{AkAL}.
Those  consist in defining a curve on the group
manifold which passes through both the background field
$\bar{\Sigma}$ and the background field with the fluctuation
$\Sigma$. One then introduces a one-parameter family of
group elements  $\Sigma(x,t)$ such that
\begin{equation}
\Sigma(x,0)= \bar{\Sigma} (x) = \xi^2 (x)  \ , \qquad 
\Sigma(x,1)= \Sigma (x) = \xi(x)  \exp( i \pi/f_\pi)  \xi(x) \ .
\end{equation}

A $t$-dependent action $S(t)$ is then defined.
The action incorporating the fluctuations to all orders can be written as
\begin{equation}
S(1) - S(0) = \int_{0} ^{1} d t \, \frac {\partial S(t)}{\partial t}
\ .
\label{taction}
\end{equation}

By introducing an exponential parametrization of  the group
element  $\Sigma(x,t)= 
\xi  \exp( i t \pi/f_\pi)  \xi$, the r.h.s. of (\ref{taction})
can be straightforwardly computed. It is enough for our purposes
to keep only the  quadratic term in $\pi$, which can be
expressed as an integral in four dimensions.
Details of this computation, as well as $T=0$ one-loop corrections
to  the WZW action can be found in Ref. \cite{AkAL}.

The  quadratic term in $\pi$ coming from the expansion
of ${\cal L}_2$ has already been computed in   Sec.  \ref{sec3}.
The  quadratic term
in the fluctuation arising from ${\cal L}_{WZ}$ is \cite{AkAL}
\begin{eqnarray}
{\cal L}_{WZ} ^{(2)} & = & - \frac{i N_c}{48 \pi^2 f_\pi ^2}
\epsilon^{\mu \nu \alpha \beta} 
Tr \Big ( (\pi d_\mu \pi -  d_\mu \pi \pi) \{\Gamma_{\nu
\alpha}, \Delta_{\beta} \} \Big. \\
& + & (\pi \Delta_\mu d_\nu \pi -  d_\nu \pi \Delta_\mu \pi)
(\Gamma_{\alpha \beta} - 2 \Delta_\alpha \Delta_\beta) 
\nonumber \\
&-&\left. \frac 18 \pi^2 [\Gamma_{\mu \nu}- 2 \Delta_\mu \Delta_\nu
, F^- _{\alpha \beta}]
- \frac 12 \pi \Delta_\mu \pi \{F_{\nu \alpha}^-, \Delta_\beta \}
\right) \nonumber \ ,
\end{eqnarray}
where the covariant derivative $d_\mu$ and the fields $\Gamma_{\mu \nu}$, 
$\Delta_\mu$ and  $F^- _{\mu \nu}$ have already been defined in Sec.
\ref{sec3}.

It is possible to express the complete quadratic term as 
\begin{equation}
\label{WZnonl}
{\cal L}_2^{(2)} + {\cal L}_{WZ}^{(2)} = \frac 12 \left( \left(\partial_\mu \pi^a +
(\Gamma_\mu ^{ab} +S_\mu ^{ab} ) \pi^b\right)^2 - \pi^a ( \sigma_0^{ab}
+\sigma_{WZ}^{ab} ) \pi^b 
+ \pi^a  S_\mu ^{ac}S_\mu ^{cb}   \pi^b \right) \ ,
\end{equation}
where
\begin{mathletters}
\begin{eqnarray}
\Gamma_\mu ^{ab} & = & - \frac 12\, Tr \left( [ \lambda^a, \lambda^b] 
\Gamma_\mu \right) \ , \\
\sigma_0 ^{ab} & = & \frac 12 \, Tr \left( [\lambda^a, \Delta_\mu] [\lambda^b, \Delta_\mu]
\right) \ , \\
S ^{\mu ab} & = & \frac{i N_c}{48 \pi^2 f_\pi ^2} \epsilon ^{\mu \nu \alpha \beta}
 Tr \left( [\lambda^a, \lambda^b] \{\Gamma_{\nu
\alpha}, \Delta_{\beta}\}
- (\lambda^a \Delta_\nu \lambda^b - \lambda^b  \Delta_\nu \lambda^a)
(\Gamma_{\alpha \beta} - 2 \Delta_\alpha \Delta_\beta) \right) \ ,
\\
\sigma_{WZ} ^{ab} & =&  \frac{i N_c}{96 \pi^2 f_\pi ^2} \epsilon^{\mu \nu \alpha \beta}
Tr \left( \frac 14 \{\lambda^a, \lambda^b\} [F^- _{\mu \nu}, \Gamma_{\alpha \beta}
- 2 \Delta_\alpha \Delta_\beta]  \right. \\
& & -  \left. (\lambda^a \Delta_\mu \lambda^b + \lambda^b  \Delta_\mu \lambda^a)
\{\Delta_\nu, F^-_{\alpha \beta}\} \right) \ .
\nonumber
\end{eqnarray} 
\end{mathletters}

Notice that both $S _{\mu}$ and $\sigma_{WZ}$ transform covariantly
under a chiral transformation.
It should also be remarked that there is not a quadratic term in $S_{\mu}$
in (\ref{WZnonl}). This is natural, since all the corrections coming from the
Wess-Zumino lagrangian are proportional to 
$ \epsilon ^{\mu \nu \alpha \beta}$.

In the special case where the external sources vanish $F^L_\mu=F^R_\mu =0$,
then $\Gamma_\mu = V_\mu$ and $\Delta_\mu = A_\mu$. In this case
 $\sigma_{WZ}=0$, and  
using the identity (\ref{useid}) for vanishing external sources one can express
\begin{equation}
S ^{\mu ab}  = - \frac{i N_c}{12 \pi^2 f_\pi ^2} \epsilon ^{\mu \nu \alpha \beta}
 Tr \left( [\lambda^a, \lambda^b] A_\nu A_\alpha A_{\beta}
- (\lambda^a A_\nu \lambda^b - \lambda^b  A_\nu \lambda^a)
 A_\alpha A_\beta \right)  \ .
\end{equation}

We will only compute thermal corrections for soft background fields
when there is only one anomalous vertex and one or several
non-anomalous ones. Thermal corrections with several anomalous
vertices are suppressed in chiral perturbation theory.

Thermal amplitudes involving only one vertex $\sigma_{WZ}$ do not produce
any one-loop correction in the hard thermal approximation. 
The only Feynman diagram that in principle could give a hard thermal correction
is a tadpole-like diagram proportional to $\sigma_{WZ}^{aa}$, where the
$SU(N)$  index is summed. However, it can be checked that  $\sigma_{WZ}^{aa}=0$.
This is the reason why the coefficient of the Wess-Zumino lagrangian does
not receive one-loop corrections, both at $T=0$ or at finite $T$. This
result could be expected since that coefficient  is fixed by topology \cite{WZW}.

Let us now evaluate a two-point function with one anomalous vertex
$S_\mu$  and one non-anomalous one $\Gamma_\mu$.
Due to the particular form of the lagrangian
one gets the same momentum dependence 
than in (\ref{twoVp}). Therefore, in the situation where
the external momenta is soft, the  same leading thermal correction is extracted. Thus 
\begin{equation}
\frac{T^2}{12} \int {\frac{d ^4 P}{(2 \pi)^4}
\, S_\mu ^{ab} (P)  \delta \Pi^{\mu \nu}(P) \Gamma_\nu ^{ab}(-P)}
\label{twoSG}
\end{equation}

There are also other amplitudes with one anomalous vertex $S_\mu$ and  $n$
non-anomalous ones  $\Gamma_\mu$ which give a non-trivial contribution to the 
thermal effective action. Instead of analyzing those Feynman diagrams
a symmetry argument will be used to obtain the complete thermal
effective action, as it was done in the previous sections.

By going from momentum to coordinate space, Eq. (\ref{twoSG}) can be
expressed as
\begin{equation}
\frac{T^2}{12} \int d^4 x\, d^4 y \int \frac {d \Omega_q}{4 \pi}
\left(\partial_\mu \Gamma_\alpha^{ab} - \partial_\alpha \Gamma_\mu^{ab}
\right)  <x | \frac{Q^\alpha Q^{\beta}}{-(Q \cdot
\partial)^2} |y> \left(\partial_\mu S_\beta^{ab} - \partial_\beta
 S_\mu^{ab}  \right) \ .
\label{2act1}
\end{equation}

Using the $SU(N)$ relation
\begin{equation}
\sum_{a=1} ^{N^2 -1} Tr(\lambda^a A) Tr(\lambda^a B) =
2 \,Tr (AB) - \frac{2}{N}\, TrA \,TrB \ ,
\end{equation}
Eq. (\ref{2act1}) becomes
\begin{equation}
\frac{T^2}{12}  \int d^4 x \, d^4 y \int \frac {d \Omega_q}{4 \pi}
Tr  \left( \left(\partial_\mu \Gamma_\alpha - \partial_\alpha \Gamma_\mu
\right)  <x | \frac{Q^\alpha Q^{\beta}}{-(Q \cdot
\partial)^2} |y> \left(\partial_\mu S_\beta - \partial_\beta
 S_\mu  \right) \right) \ ,
\label{2act2}
\end{equation}
where $S_\mu$ is defined by
\begin{eqnarray}
S^\mu & = & - \frac{i N_c}{48 \pi^2 f_\pi ^2}  \epsilon ^{\mu \nu \alpha \beta}
\sum_{a=1} ^{N^2 -1} \Big(
\left[ [ \lambda^a, \{\Gamma_{\nu
\alpha}, \Delta_{\beta}\} ] ,   \lambda^a \right]
\Big. \\
& &  - \Big.
\left[\Delta_\nu \lambda^a 
(\Gamma_{\alpha \beta} - 2 \Delta_\alpha \Delta_\beta) -
(\Gamma_{\alpha \beta} - 2 \Delta_\alpha \Delta_\beta) \lambda^a \Delta_\nu ,
\lambda^a \right] \Big ) \ .
\nonumber
\end{eqnarray}

To make the whole expression (\ref{2act2}) invariant under a chiral transformation
it is enough to substitute the ordinary derivatives $\partial_\mu$ by 
covariant ones,  $d_\mu$. Then one  gets

\begin{equation}
\frac{T^2}{12}  \int d^4 x \, d^4 y \int \frac {d \Omega_q}{4 \pi}
Tr \left(\Gamma_{\mu \alpha} 
  <x | \frac{Q^\alpha Q^{\beta}}{-(Q \cdot
d)^2} |y> \left( d_\mu S_\beta - d_\beta
 S_\mu \right) \right) \ ,
\label{2act3}
\end{equation}
which gives the thermal effective action for soft background fields 
involving only one anomalous vertex.

From this effective action one could get 
the thermal corrections to several anomalous decays,
 such as $\pi^0 \rightarrow \gamma \gamma$
\cite{PT2}.

\section{CONCLUSIONS}
\label{sec5}

In a non-Abelian gauge theory the one-loop thermal correction to
the polarization tensor is easily computed when the external momenta
of the gluon is soft,  $P \sim g T \ll T$. Recquiring these thermal  effects
be gauge invariant allows to obtain the one-loop thermal effective action
for soft modes. 
This effective action describes thermal amplitudes with $n$ external 
soft gluon legs, with $n=2, \ldots, \infty$, the so called HTL's, 
and represents a thermal mass for the longitudinal gauge degrees of
freedom.

 In the framework of chiral perturbation theory also a symmetry
argument allows to derive the complete hard thermal loop effects
from the knowledge of the two-point functions. This is the idea
that has been exploited in this article. 

The same thermal amplitudes, the HTL's,  appear in a 
gauge field theory and in a chiral lagrangian. However, 
it should be clear that the physics associated to the HTL's is quite 
different in the two different models, as it has been already emphasized.

In a gauge theory HTL's are thermal amplitudes which are as important
as the tree amplitudes for soft momenta, and thus they have to be resummed
into effective vertices and propagators. The gauge coupling constant 
$g$, which at high $T$ is $g \ll 1$,
 allows to identify neatly the two different energy scales $T$ and 
$g T$ as two relevant physical scales of the theory.

In the chiral lagrangians that 
have been considered here HTL's only represent corrections to 
$\pi -\pi$ scattering of order
$T^2/f_\pi ^2$, and they do not have to be resummed.
It is actually the dimensionless quantity $T^2/f_\pi ^2$, which is assumed
to be $T^2/f_\pi ^2 \ll 1$, the expansion parameter for the thermal
corrections in chiral perturbation theory.  It should be stressed that 
this parameter allows also to
separate neatly the physical scales {\it hard} ($\sim T$) or {\it soft} ($\sim
 \sqrt{(T^2/f_\pi ^2)} T$) in  a chiral theory.

A power counting analysis could be  established to obtain
the leading thermal corrections in an arbitrary Feynman diagram
in chiral perturbation theory. This counting is essential to get correctly
the two-loop thermal corrections arising from ${\cal L}_2$, and one-loop
thermal corrections from  ${\cal L}_4$. 
It is then a remaining open question whether one could study those
thermal effects with the same kind of techniques that have been used
in this article.

\vfill
\eject

{\bf Acknowledgments:}

I want to thank F. Brandt, V. Koulovassilopoulos 
and J. Taron for useful discussions, and to R.~Jackiw for pointing out the relevance
of some references. 
This work has been supported  by funds provided by the
CICYT contract AEN95-0590, and by the CIRIT contact GRQ93-1047.

\end{document}